\documentclass[runningheads]{llncs}
\usepackage{graphicx}
\usepackage{todonotes}
\usepackage[export]{adjustbox}
\newcommand{\xiaozhou}[1]{\textcolor{black}{#1}}

\begin{document}
\title{OSS PESTO: An Open Source Software Project Evaluation and Selection TOol}
%
%

\author{Xiaozhou Li \and
Sergio Moreschini}

\authorrunning{X. Li and S. Moreschini}

%
\institute{Tampere University\\
Kalevantie 4, 33100, Tampere, Finland \\
\email{\{xiaozhou.li, sergio.moreschini\}@tuni.fi}}

\maketitle              
\begin{abstract}


Open source software (OSS), playing an increasingly critical role nowadays, has been commonly adopted and integrated in various software products. For many practitioners, selecting and adopting suitable OSS can help them greatly. Though many studies have been conducted on proposing OSS evaluation and selection models, a limited number are followed and used in the industry. Meanwhile, many existing OSS evaluation tools, though providing valuable details, fall short on offering intuitive suggestions in terms of framework-supported evaluation factors. Towards filling the gap, we propose an Open Source Software Project Evaluation and Selection TOol (OSS PESTO). Targeting OSS on Github, the largest OSS source code host, it facilitates the evaluation practice by enabling practitioners to compare candidates therein in terms of selected OSS evaluation models. It also allows in-time Github data collection and customized evaluation that enriches its effectiveness and ease of use.  


\end{abstract}
\section{Introduction}


During the last two decades, open source software (OSS) has been flourishing with such trend continuing~\cite{Kilamo2020} and nowadays, OSS is adopted by the vast majority of IT companies~\cite{Lenarduzzi2019}\cite{Lenarduzzi2020SEAA}.  On Github, the largest OSS source code host, more than 60 million users\footnote{https://github.com/search?q=type:user\&type=Users} have participated in over 100 million open source projects\footnote{https://github.com/search?q=is:public}, among which many have been widely adopted by users and companies. However, due to such a large number of candidates, for many practitioners, selecting a suitable OSS product or library is difficult, especially when the relevant information are not explicitly provided \cite{Lenarduzzi2020SEAA}. 

To support the OSS evaluation and selection practice, many studies provide models and frameworks as guidance \cite{duijnhouwer2003capgemini,QSOSTool,Davide_etal_2007,Wasserman_etal_2006,wasserman2017osspal}. During the last two decades, 35 models are proposed with checklists, measures or both provided \cite{Lenarduzzi2020SEAA}. They all work similarly with a process of ``candidate software identification - factor evaluation - scoring". In addition, many tools are designed and proposed to facilitate such practice \cite{QSOSTool,Davide_etal_2007,ROSEN,wasserman2017osspal}. However, many tools are rigidly designed and allow limited customization. Meanwhile, among those proposed, only two are properly maintained with the majority being not available any more \cite{Lenarduzzi2020SEAA}.

As the largest OSS host and platform, Github is a valuable channel restoring and presenting OSS related information. Retrospective analysis of Github repositories, based on the abundant information on code, developers, organizations and activities within, can yield valuable insights into the evolution and growth of OSS and facilitate decision making processes \cite{munaiah2017curating}. Many studies have used such data source and conduct research on various OSS related perspectives \cite{kalliamvakou2014promises,lima2014coding,tsay2014influence}. However, the approaches towards using Github data for OSS evaluation are limited, let alone tools to support such practice. 

Herein, we propose OSS PESTO, an open source tool facilitating OSS evaluation and selection. Comparatively, besides being fully open sourced and free to use, OSS PESTO has the following advantages: 1) it allows users to update in-time Github data; 2) it allows users to customize evaluation with models or preference; 3) it allows users to save data locally and to use it without network connection; 4) it is always accessible as maintained in Github repository. It shall largely help the practitioners to compare and evaluate OSS candidates freely, timely and efficiently.

The remainder of this paper is organized as follows. Section 2 introduces the related work on OSS evaluation tools. Section 3 presents OSS PESTO with details. Section 4 presents an experiment validating its applicability. Section 5 concludes the article.

\section{Related Work}

The Open Source Maturity Model (OSMM) is the first proposed model and open standard that aims for such purpose \cite{duijnhouwer2003capgemini}. Guided by OSMM, the practitioners will evaluate OSS by its maturity of each aspect, weight each aspect with importance, and compute its overall maturity by the weighted sum. 
Compared to OSMM, the Open Business Readiness Rating (OpenBRR) is an OSS evaluation method with more indicators, the idea of target uses and the customized evaluation \cite{Wasserman_etal_2006}. The method provides an index applicable to all OSS development initiatives. Its main limits are related to the incompatibility of the requirements between different targets and to the difficulty of choosing the proper reference application for some projects. 
Similarly, a number of evaluation models are proposed, for example, Qualification and Selection of Open Source Software (QSOS) \cite{QSOSTool}, OpenBQR \cite{Davide_etal_2007}, OSSPAL \cite{wasserman2017osspal}, which provide enhanced guidance and methodological support. 

QSOS tool is designed to support the QSOS model which aims to qualify, select and compare OSS products \cite{QSOSTool}. However, the rigidness of compulsory Identity Card setting and all criteria inclusion is commonly seen as its limitation.
OpenQBR \cite{Davide_etal_2007} requires specification on factor importance before the assessment of the project. Compared to the QSOS tool, OpenQBR is more elastics as not require to evaluate factors which are not relevant to the specific project. 

OSS-PAL \cite{wasserman2017osspal}, though similar to QSOS, aims to partially automate the evaluation of the projects. Despite the appealing goal of the project, it fails to provide the automated data collection function. 
Other works investigated the availability of the information on online portals~\cite{Sbai2018}\cite{Li2021}, but they did not provide tools for collecting or aggregating data.

In addition, many other tools are available over time but have been discontinued, including real-time OpenSSL execution monitoring system (ROSEN) \cite{ROSEN}, RAP TOOL \cite{RAP}, SQO-OSS \cite{SQO}, OMM Tool \cite{OMM}, T-Doc Tool \cite{TDOC}, QualiPSo Trustworthiness Checklist \cite{qualipso}, MOSST \cite{MOSST} and OP2A Checklist \cite{OPA2} and other checklist included in marketing models for OSS~\cite{Lenarduzzi2011}\cite{delBianco2012}. 

With OSS PESTO, we aim to overcome some of the most common drawbacks of all of these tools, such as, the focus on specific factors, the evaluation of factors before adding a weight function or the lack of control for both internal and external product quality.

\section{OSS PESTO}




We implement OSS PESTO\footnote{Source code: https://github.com/clowee/OSS-PESTO} by following the commonly acknowledged OSS evaluation process summarized from previous studies \cite{Lenarduzzi2020SEAA}. It shall contain the following main activities: 1) identify the OSS candidates; 2) elicit a list of factors that need to be evaluated and the according metrics that measure such factors; 3) provide scores or selection recommendation as evaluation output. 

In addition, in order to use the latest Github data to evaluate OSS, we integrate a data crawler module in OSS PESTO. It enables the users to crawl the required repository and activities information of any existing OSS projects. Additionally, it also allows them to crawl the data of a list of projects based on the selected range of stars. Furthermore, OSS PESTO allows users to customize evaluation factors based on the selection of models and/or their personal preferences. 

\begin{figure*}[h]
\vspace{-0.1in}
\begin{center}
\includegraphics[width=\textwidth]{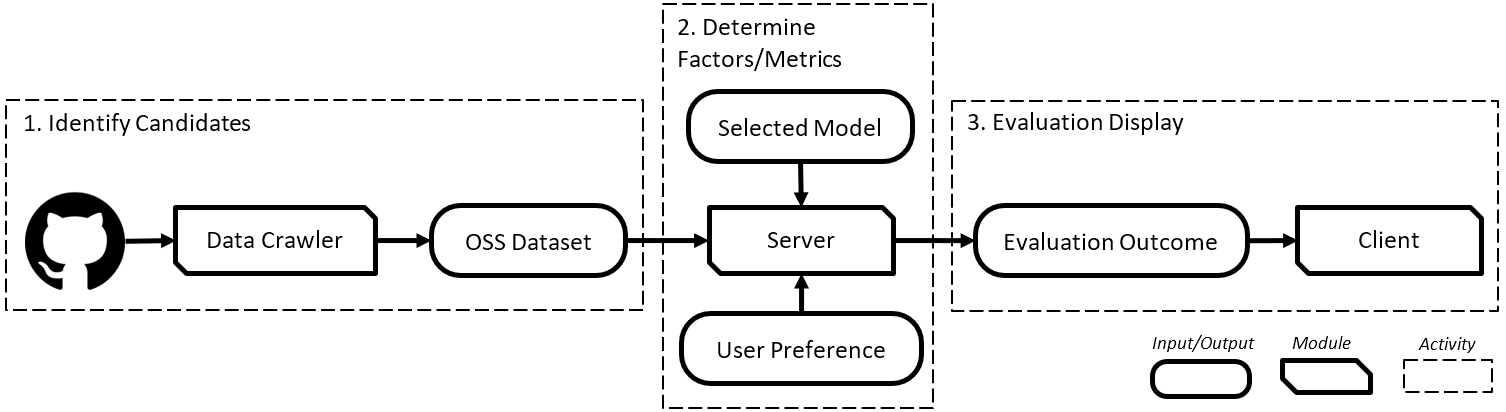}
\caption{OSS PESTO Framework}
\label{fig:frame}
\end{center}
\vspace{-0.3in}
\end{figure*}

Shown in Fig. \ref{fig:frame}, OSS PESTO contains three individual modules as follows.

\begin{itemize}
    \item \textbf{Data Crawler}: The data crawler module contains a set of Python scripts that extract Github repository data via Github APIs\footnote{https://docs.github.com/en/graphql; https://docs.github.com/en/rest}. It enables the users to select the candidate OSS and extract the according data. 
    \item \textbf{Server}: The server side is implemented by ReactJS\footnote{https://reactjs.org/} while database with MongoDB\footnote{https://www.mongodb.com/}. The evalaution model is described with the config.json file, which can be altered with users' preference of evaluation factors.
    \item \textbf{Client}: The client side is also implemented by ReactJS. It mainly displays the candidate OSS projects with the selected attributes/factors shown. It also shows the results of candidate comparison which facilitates OSS evaluation and selection.
\end{itemize}

Fig. \ref{fig:frame} also shows the activities of utilizing OSS PESTO to evaluate candidate OSS projects as follows.

\begin{itemize}
    \item \textbf{Step 1}. identifies the OSS candidates by running the data crawler module to extract the according dataset. 
    \item \textbf{Step 2}. select the evaluation model, configure evaluation preference, and run the server module.
    \item \textbf{Step 3}. run the client module and compare the OSS candidates by the selected factors.
\end{itemize}

\xiaozhou{The crawled data is saved locally in A comma-separated values (CSV) file with each row containing the values of an individual OSS candidate. To be noted, the required data can be selectively crawled according to the users, who determine which metrics are the important ones when evaluating particular aspects of OSS. Such selection of data can be guided by the evaluation model chosen by the evaluator. For example, when selecting only the most popular OSS, the numbers of stars, watches, and download are the ones to be crawled.}

\begin{figure*}[h]
\vspace{-0.1in}
\begin{center}
\includegraphics[width=\textwidth]{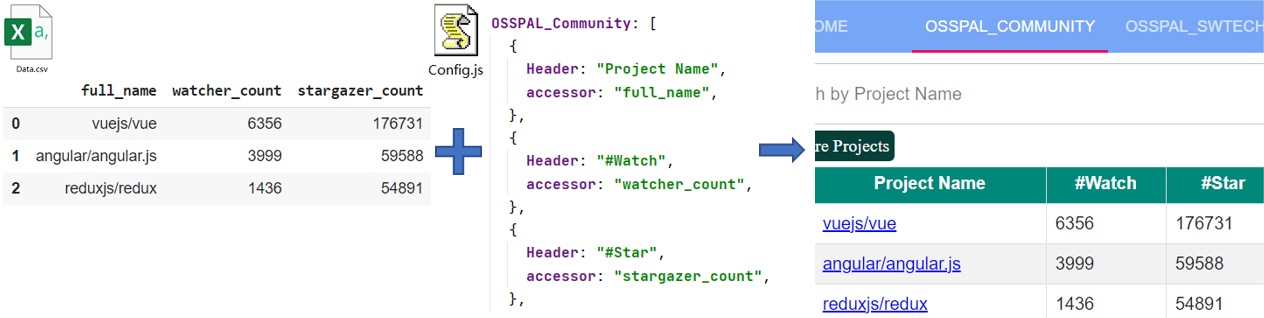}
\caption{An example of Configuration File}
\label{fig:config}
\end{center}
\vspace{-0.3in}
\end{figure*}

\xiaozhou{Furthermore, the configuration file is a Javascript file mapping the category tabs displayed by the client and the data features/metrics that are selected to evaluate the according categories. Shown in \ref{fig:config} is an example of how a configuration file works. By editing the configuration file, the users can customize their selection of metrics, the evaluation categories and the links in between. For example, if the user chooses to focus on the popularity of OSS and uses the number of watches as the metric for it, the according piece of code \textit{\{ Header: ``\#Watch", accessor: ``watcher\_count" \}} shall be added to the ``Popularity" tab block.}

\section{Experiment Showcase}


In order to validate the applicability of OSS PESTO, we conduct a series of experiments, including the testing of all three modules. The testing scenario is to evaluate and compare three JavaScript frameworks, i.e., Angular\footnote{https://angular.io/}, Redux\footnote{https://redux.js.org/} and Vue\footnote{https://vuejs.org/} using the OSSPAL model \cite{wasserman2017osspal}. The evaluation categories include ``Community", ``Support", ``Operational Software Characteristics", ``Documentation", ``Software Technology Attributes", ``Functionality" and ``Development Process". Herein, we focus on the ``Community", ``Support" and ``Software Technology Attributes" aspects, which can be well demonstrated by the obtained data.

To start crawling the Github data, given the user's Github personal token and the target OSS candidates as input, the data crawler module can be ran individually and continuously. Towards the stated objective, the crawling process takes within two minutes. When the data is ready, we prepare the config.json by selecting the target metrics that are valuable towards evaluating each factors of the candidates. For each of the selected factors, the according metrics are as follows.

\begin{itemize}
    \item Community: number of watches, number of stars, age, average issue active time, average issue comments, number of pull requests, and number of issue raiser.
    \item Support: average issue closed time, number of contributor, organization issue raiser.
    \item Software Technology Attributes: number of open issues, number of dependence.
\end{itemize}

\begin{figure*}[h]
\vspace{-0.1in}
\begin{center}
\includegraphics[width=\textwidth]{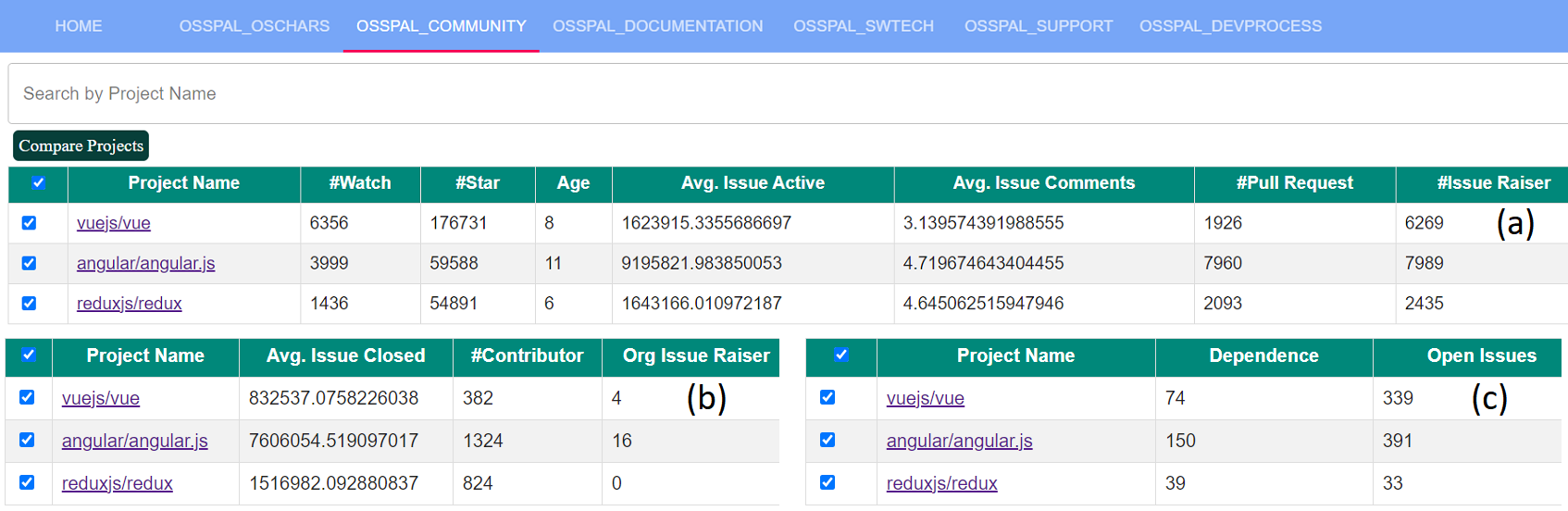}
\caption{Experiment Results Demonstration}
\label{fig:result1}
\end{center}
\vspace{-0.3in}
\end{figure*}

Thereafter, when running both the server and the client, the comparison result is shown in Fig. \ref{fig:result1}.

Based on such comparison, we can easily observe that despite not being the oldest community, Vue is more popular than the other two candidates in terms of watches and stars. However, these three communities are active in different ways, as Angular has more comments on issues, pull requests, and different issue raisers while the others are more responsive to issues (shown in Fig.\ref{fig:result1} (a)). Regarding support, Angular has a much larger contributor group and organizational issue raiser for support, while on software technology attribute aspect, Redux has much less dependence and open issues (shown in Fig.\ref{fig:result1} (b) and (c)). 

When adopting a different evaluation model, it is possible that by taking into account particular overseen metrics, the user obtains new insights regarding the selected candidates. For example, the SQO-OSS model \cite{SQO} sees ``Growth in active developers" as a metric to evaluate the ``Developer base" category, when OSSPAL has not such category. However, due to the fact that same dataset is used for all potential models, it is hardly possible to have opposite comparative evaluation result for the same category from different models.

\section{Conclusion}




This paper presents OSS PESTO, an open source software project evaluation and selection tool, to support the practitioners' need towards OSS evaluation and selection. This tool provides a Github-repository-data-oriented, easy-to-maintain, customization-friendly solution. It shall benefit the practitioners in both industry and academia in terms of the different focuses on either the OSS projects or the OSS evaluation models respectively. 

However, the current version of this tool can certainly be improved in the following ways. \xiaozhou{Firstly, OSS PESTO has not yet supported the practitioners' selection of OSS candidates at the identification phase in terms of their target functionalities. As the accessible data obtained from Github does not provide explicit information regarding the main features of the OSS, such candidate selection cannot be automated via direct identification. A potential solution is to apply natural language processing (NLP) techniques to identify and summarize such main features from the description and Readme text of the projects. Such a feature shall be implemented in our future work.}

\xiaozhou{Furthermore, the current version only utilizes limited amount of the attributes provided by the Github API. For many such attributes, the explicit mappings towards particular OSS evaluation categories are not verified. For example, the number of OSS downloads can be seen as a metric for its popularity. However, unless a particular user insists it being a critical evaluation criterion for his/her customized evaluation model, such value can be ignored when it does not contribute to any pre-defined evaluation categories. Nonetheless, the inclusion of more data features shall be taken into account in the future work. However, it should be noted such work can result in the exhaustion of Github API query limit, as some values (e.g., issues) can only be obtained via looping enumerated results. }

In addition, more features, in terms of the ease of use perspective of the tool, shall be also considered. For example, a graphic user interface is needed for the data crawler module which can also be integrated to the server side. Furthermore, the data from Github has its limitation on reflecting certain aspects of OSS. For example, the development process of the projects cannot be easily accessed externally, except for the number of releases and the release pace. Thus, in order to improve the potential scope of this tool, more data sources are required with more techniques required to process possibly unstructured data as well. Meanwhile, more practical features, such as, exporting the evaluation results, adding weight to different factors, editor of configure files, and model customization interface, are also required. 

Our future work shall focus on integrating the modules and enhancing the overall quality of the tool according to the above mentioned limitation. It is also important to investigate the ways of evaluating individual OSS by providing unified quantified results. In addition, we shall systematically investigate the availability of data from multiple sources that could be used to support OSS evaluation. 

\bibliographystyle{splncs04}
\bibliography{cite}

\begin{thebibliography}{10}
\providecommand{\url}[1]{\texttt{#1}}
\providecommand{\urlprefix}{URL }
\providecommand{\doi}[1]{https://doi.org/#1}

\bibitem{OPA2}
Benlian, A., Hess, T.: Comparing the relative importance of evaluation criteria
  in proprietary and open-source enterprise application software selection--a
  conjoint study of erp and office systems. Information Systems Journal
  \textbf{21}(6),  503--525 (2011)

\bibitem{delBianco2012}
del Bianco, V., Lavazza, L., Lenarduzzi, V., Morasca, S., Taibi, D., Tosi, D.:
  A study on oss marketing and communication strategies. In: Open Source
  Systems: Long-Term Sustainability. pp. 338--343. Springer Berlin Heidelberg,
  Berlin, Heidelberg (2012)

\bibitem{qualipso}
del Bianco, V., Lavazza, L., Morasca, S., Taibi, D., Tosi, D.: The qualispo
  approach to oss product quality evaluation. In: Proceedings of the 3rd
  International Workshop on Emerging Trends in Free/Libre/Open Source Software
  Research and Development. pp. 23--28 (2010)

\bibitem{ROSEN}
Choi, S.j., Kang, Y.h., Lee, G.s.: A security evaluation and testing
  methodology for open source software embedded information security system.
  In: International Conference on Computational Science and Its Applications.
  pp. 215--224. Springer (2005)

\bibitem{OMM}
Del~Bianco, V., Lavazza, L., Morasca, S., Taibi, D.: Quality of open source
  software: the qualipso trustworthiness model. In: IFIP International
  Conference on Open Source Systems. pp. 199--212. Springer (2009)

\bibitem{MOSST}
Del~Bianco, V., Lavazza, L., Morasca, S., Taibi, D., Tosi, D.: A survey on the
  importance of some economic factors in the adoption of open source software.
  In: Software Engineering Research, Management and Applications 2010, pp.
  151--162. Springer (2010)

\bibitem{duijnhouwer2003capgemini}
Duijnhouwer, F.W., Widdows, C.: Capgemini expert letter open source maturity
  model. Capgemini pp. 1--18 (2003)

\bibitem{RAP}
Immonen, A., Palviainen, M.: Trustworthiness evaluation and testing of open
  source components. In: Seventh International Conference on Quality Software
  (QSIC 2007). pp. 316--321. IEEE (2007)

\bibitem{kalliamvakou2014promises}
Kalliamvakou, E., Gousios, G., Blincoe, K., Singer, L., German, D.M., Damian,
  D.: The promises and perils of mining github. In: Proceedings of the 11th
  working conference on mining software repositories. pp. 92--101 (2014)

\bibitem{Kilamo2020}
Kilamo, T., Lenarduzzi, V., Ahoniemi, T., Jaaksi, A., Rahikkala, J., Mikkonen,
  T.: How the cathedral embraced the bazaar, and the bazaar became a cathedral.
  In: Ivanov, V., Kruglov, A., Masyagin, S., Sillitti, A., Succi, G. (eds.)
  Open Source Systems. pp. 141--147. Springer International Publishing, Cham
  (2020)

\bibitem{Lenarduzzi2011}
Lenarduzzi, V.: Towards a marketing strategy for open source software. In:
  Proceedings of the 12th International Conference on Product Focused Software
  Development and Process Improvement. p. 31–33. Profes '11, Association for
  Computing Machinery, New York, NY, USA (2011). \doi{10.1145/2181101.2181109}

\bibitem{Lenarduzzi2020SEAA}
Lenarduzzi, V., Taibi, D., Tosi, D., Lavazza, L., Morasca, S.: Open source
  software evaluation, selection, and adoption: a systematic literature review.
  In: 2020 46th Euromicro Conference on Software Engineering and Advanced
  Applications (SEAA). pp. 437--444 (2020). \doi{10.1109/SEAA51224.2020.00076}

\bibitem{Lenarduzzi2019}
Lenarduzzi, V., Tosi, D., Lavazza, L., Morasca, S.: Why do developers adopt
  open source software? past, present and future. In: Open Source Systems. pp.
  104--115. Springer International Publishing, Cham (2019)

\bibitem{Li2021}
Li, X., Moreschini, S., Zhang, Z., Taibi, D.: Exploring factors and measures to
  select open source software. In: Arxiv (2021)

\bibitem{lima2014coding}
Lima, A., Rossi, L., Musolesi, M.: Coding together at scale: Github as a
  collaborative social network. In: Proceedings of the International AAAI
  Conference on Web and Social Media. vol.~8 (2014)

\bibitem{TDOC}
Morasca, S., Taibi, D., Tosi, D.: T-doc: A tool for the automatic generation of
  testing documentation for oss products. In: IFIP International Conference on
  Open Source Systems. pp. 200--213. Springer (2010)

\bibitem{munaiah2017curating}
Munaiah, N., Kroh, S., Cabrey, C., Nagappan, M.: Curating github for engineered
  software projects. Empirical Software Engineering  \textbf{22}(6),
  3219--3253 (2017)

\bibitem{QSOSTool}
Origin, A.: Method for qualification and selection of open source software
  (qsos). \url{http://www.qsos.org} (Accessed: 2021-01-22)

\bibitem{SQO}
Samoladas, I., Gousios, G., Spinellis, D., Stamelos, I.: The sqo-oss quality
  model: measurement based open source software evaluation. In: IFIP
  international conference on open source systems. pp. 237--248. Springer
  (2008)

\bibitem{Sbai2018}
Sbai, N., Lenarduzzi, V., Taibi, D., Sassi, S.B., Ghezala, H.H.B.: Exploring
  information from oss repositories and platforms to support oss selection
  decisions. Information and Software Technology  \textbf{104},  104--108
  (2018). \doi{https://doi.org/10.1016/j.infsof.2018.07.009},
  \url{https://www.sciencedirect.com/science/article/pii/S0950584918301526}

\bibitem{Davide_etal_2007}
Taibi, D., Lavazza, L., Morasca, S.: Openbqr: a framework for the assessment of
  oss. In: Feller, J., Fitzgerald, B., Scacchi, W., Sillitti, A. (eds.) Open
  Source Development, Adoption and Innovation. pp. 173--186. Springer US,
  Boston, MA (2007)

\bibitem{tsay2014influence}
Tsay, J., Dabbish, L., Herbsleb, J.: Influence of social and technical factors
  for evaluating contribution in github. In: Proceedings of the 36th
  international conference on Software engineering. pp. 356--366 (2014)

\bibitem{wasserman2017osspal}
Wasserman, A.I., Guo, X., McMillian, B., Qian, K., Wei, M.Y., Xu, Q.: Osspal:
  finding and evaluating open source software. In: IFIP International
  Conference on Open Source Systems. pp. 193--203. Springer, Cham (2017)

\bibitem{Wasserman_etal_2006}
Wasserman, A.I., Pal, M., Chan, C.: The business readiness rating: a framework
  for evaluating open source technical report  (2006)

\end{thebibliography}
\end{document}